\documentclass[10pt,aps,twocolumn, prl, footinbib,superscriptaddress]{revtex4-1}
\usepackage{mathrsfs}
\usepackage{graphicx}% Include figure files
\usepackage{dcolumn}% Align table columns on decimal point
\usepackage{bm}% bold math
\usepackage{amsmath,amssymb}

\begin{document}

\title{Controlled Transport between Fermi Superfluids Through a Quantum Point Contact}
\author{Juan Yao}
\affiliation{Institute for Advanced Study, Tsinghua University,
Beijing, 100084, China}

\author{Boyang Liu}
\email{byliu@bjut.edu.cn}
\affiliation{Institute of Theoretical Physics, Beijing University of Technology, Beijing, 100124, China}

\author{Mingyuan Sun}
\affiliation{Institute for Advanced Study, Tsinghua University, Beijing, 100084, China}

\author{Hui Zhai}
\affiliation{Institute for Advanced Study, Tsinghua University, Beijing, 100084, China}
\affiliation{Collaborative Innovation Center of Quantum Matter, Beijing, 100084, China}

\date{\today}

\begin{abstract}

Recent advances in experimental techniques allow one to create a quantum point contact between two Fermi superfluids in cold atomic gases with a tunable transmission coefficient. In this Letter we propose that three distinct behaviors of charge transports between two Fermi superfluids can be realized in this single setup, which are the multiple Andreev reflection, the self-trapping and the Josephson oscillation. We investigate the dynamics of atom number difference between two reservoirs for different initial conditions and different transmission coefficients, and present a coherent picture of how the crossover between different regimes takes place. Our results can now be directly verified in current experimental system.

\end{abstract}
\maketitle

\begin{figure}[t]
\includegraphics[width=0.4\textwidth]{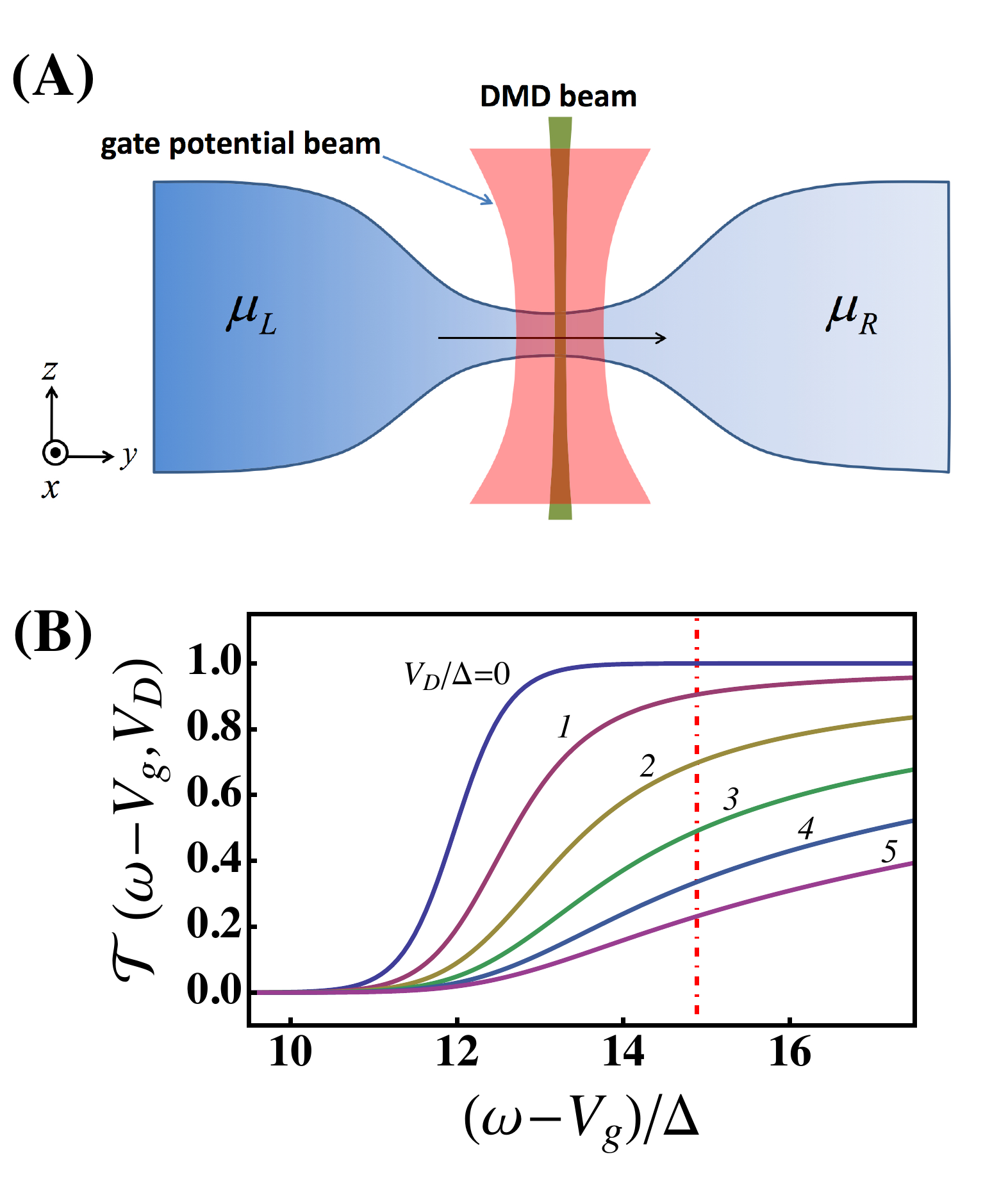}
\caption{(Color online) (A) The geometry of the experimental setup. Two reservoirs are connected by a QPC. A gate beam applied on the central regime tunes the relative energy $V_\text{g}$ between the QPC and the reservoirs. A tightly focused DMD beam tunes the tunneling amplitude. (B) The transmission coefficient $\mathcal{T}$ as a function of $\omega$ is plotted for different $V_\text{D}$. For later convenience, we have taken the bulk pairing gap $\Delta$ as the energy unit and we choose $\hbar\omega^0_x/\Delta=18$ and $\hbar\omega^0_z/\Delta=6$. }
\label{fig:model}
\end{figure}

Transport measurements are powerful tools not only for revealing fundamental properties of quantum materials in condensed matter physics, but also for constructing solid-state devices. In the past few years, transport has also become one of the frontiers in cold atom physics. Various experiments have been conducted, including particle transport \cite{Stadler,Krinner1,Valtolina,Husmann,Krinner2,Hausler}, heat transport \cite{Brantut}, and spin transport \cite{Sommer,Bardon,Koschorreck,Luciuk}. Of particular interests is the realization of the two-terminal transport measurements in cold atom setup \cite{Stadler,Krinner1,Valtolina,Husmann,Krinner2,Brantut,Hausler}. As shown in Fig. \ref{fig:model}(A), a cigar-shaped cloud is split into two reservoirs and connected by a quantum point contact (QPC) generated by high-resolution lithography \cite{Krinner1,Husmann,Krinner2}. With this setup, in the weakly interacting regime, quantized conductance of neutral matter has been first observed \cite{Krinner1}. In the strongly interacting regime, both multiple Andreev reflections between two Fermi superfluids and anomalous transport between two normal gases have been observed \cite{Husmann,Krinner2}, the later of which has attracted considerable attentions for the lack of theoretical consensus on its origin \cite{Liu2014,Glazman,Liu2017,Uchino}.

One great advantage of studying transport with cold atoms is to utilize the tunability of this system to cover different physics in different parameter regimes in a single setting, and therefore to reveal how the transition between them takes place. Here we focus on the case of the Fermi superfluid. One well known transport effect is the Josephson effect, which is a non-dissipative coherent oscillation. It has also been observed in cold atom experiments for both Bose and Fermi superfluids \cite{Albiez,Valtolina} when two superfluids are separated by a potential barrier. In contrast, dissipative transport can also take place between two fermionic superfluids. In the experiment of Ref. \cite{Husmann}, when two Fermi superfluids are connected by a QPC, the time evolution of the particle number difference between two reservoirs exhibits a non-exponential decay, which can be well explained by the multiple Andreev reflections.

Then, a question is raised naturally. For transport between two Fermi superfluids, under what condition it exhibits the Josephson effect and under what condition it exhibits the multiple Andreev reflections. In fact, the single particle transmission coefficient plays a crucial role in answering this question. A small transmission coefficient favors the Josephson effect and a large transmission coefficient favors the multiply Andreev reflections. Fortunately, recently a new technique allows one to continuously tune the transmission coefficient in the cold atom QPC setup by imprinting a mesoscopic potential or a lattice into the tunneling channel with a digital micromirror device (DMD) \cite{Hausler, DMD}. The key point of this Letter is to propose that one can use this new technology to tune the transport between two Fermi superfluids, with which one can observe and understand the crossover from the multiple Andreev reflection type transport to the self-trapping and the Josephson type transport. This will provide a unified view of these seemingly disparate transport phenomena.

\textit{Setup.} The setup of a QPC for cold atom system is schematically shown in Fig. \ref{fig:model}(A). First of all, two beams provide transverse harmonic confinements along $\hat{x}$ and $\hat{z}$ directions, respectively, whose confinement frequencies vary smoothly along the longitudinal transport direction $\hat{y}$. That is to say, in a first quantized form, the Hamiltonian for the QPC regime can be written as
\begin{equation}
\hat{H}=\frac{{\bf p}^2}{2m}+\frac{1}{2m}\omega^2_x(y)x^2+\frac{1}{2m}\omega^2_z(y)z^2+V_\text{DMD}(y),
\end{equation}
where $\omega_i(y)=\omega^0_i e^{-y^2/d^2_i}$ ($i=x,z$). Here typical values for experiment are $\hbar\omega^0_x\sim12E_F$, $\hbar\omega^0_z\sim 4E_F$, $d_x\sim3.6/k_F$ and $d_z\sim18.2/k_F$ with total number of particles $N\sim2\times10^5$. Because $d_x>d_z$, it first squeezes the system into a quasi-two-dimensional plane and then to a quasi-one-dimensional tunnel. Secondly, another beam provides a gate potential that generates a uniform energy shift $V_\text{g}$ between the regime of QPC and the reservoir. That is to say, when we consider an incoming state whose asymptotic behavior is a plane wave with momentum $k_y$, the energy conservation gives $\omega-V_\text{g}=\hbar^2 k^2_y/(2m)$. Finally, a DMD beam can generate one or a sequence of delta-function potentials inside the quasi-one-dimensional tunneling channel. For the simplest case, we first consider $V_\text{DMD}(y)=V_\text{D}\delta(y)$.

In practice, $V_\text{g}$ and $V_\text{D}$ are two parameters that can be easily tuned. By solving this QPC Hamiltonian, one can obtain the transmission coefficient $\mathcal{T}$, as shown in Fig. \ref{fig:model}(B). Without $V_\text{DMD}(y)$, the potential varies sufficiently smoothly in space such that $\mathcal{T}$ sharply jumps from zero to unity, when the incoming energy $E$ increases beyond the threshold $\hbar(\omega^0_x+\omega^0_z)/2$ and one of the tunneling channels becomes open, as shown by the $V_\text{D}=0$ curve in Fig. \ref{fig:model}(B). For finite $V_\text{D}$, $\mathcal{T}$ varies much more smoothly as $\omega$ varies. Consequently, for a fixed energy $\omega$ above the threshold, $\mathcal{T}$ continuously decreases as $V_\text{D}$ increases. In this way, we can tune the transmission coefficient in this system.

\textit{Model and Method.} First of all, we take a mean-field Hamiltonian to describe the Fermi superfluids in the left and the right reservoirs, which reads
\begin{equation}\begin{aligned}
\hat{H}_j=\sum_{\bf k \sigma}\xi_{j {\bf k}}\hat{\psi}_{j\sigma}^\dagger({\bf k})\hat{\psi}_{j\sigma}({\bf k})
-\Delta_j\hat{\psi}_{j\uparrow}({\bf k})\hat{\psi}_{j\downarrow}(-{\bf k})+\text{h.c.} ,
\end{aligned}\end{equation}
where $j=\text{L}, \text{R}$ is the reservoir index, $\sigma=\uparrow,\downarrow$ labels spin index, and $\xi_{j{\bf k}}={\bf k}^2/(2m)-\mu_j$. The parameters $\mu_j$ and $\Delta_j$ are the chemical potential and the order parameter of the $j$-th reservoir, respectively. Here, as an example, we will take $\mu/E_\text{F}=0.59$ and $\Delta/E_\text{F}=0.68$ as typical values for unitary Fermi gas.

Secondly, to study the transport behavior, we employ the non-equilibrium Keldysh formalism, for which we introduce the forward and backward  branches of the time contours and denote them by $\alpha=1,2$ after the Keldysh rotation \cite{Kamenev}. Since later we will model the QPC as a local tunneling from the left to the right reservoirs, we introduce
\begin{equation}
\hat{\psi}_{j\alpha\sigma}(\omega-\mu_j)=\int dt \hat{\psi}_{j\alpha\sigma}({\bf r}=0,t)e^{i(\omega-\mu_j) t},
\end{equation} where $\omega$ is defined as the absolute energy of the particles, and thus $\omega-\mu_j$ is the energy measured with respect to the Fermi surface of the $j$-th reservoir.
Here we define a spinor as $\hat{\Psi}(\omega)=(\hat{\Psi}_L(\omega-\mu_L), \hat{\Psi}_R(\omega-\mu_R))^T$ with
\begin{equation}\hat{\Psi}_j(\omega-\mu_j)=\begin{bmatrix}
\hat{\psi}_{j1\uparrow}(\omega-\mu_j) \\ \hat{\psi}^\dag_{j2\downarrow}(\overline{{\omega}-\mu_j}) \\ \hat{\psi}_{j2\uparrow}(\omega-\mu_j) \\\hat{\psi}^\dag_{j1\downarrow}(\overline{\omega-\mu_j})
\end{bmatrix},
\end{equation}
where $\overline{{\omega}-\mu_j}=-({\omega}-\mu_j)$ and ``$T$" denotes the transposition. The Green's function $\mathcal{G}_0=\langle \hat{\Psi}(\omega)\hat{\Psi}^\dag(\omega)\rangle$ is therefore an $8\times 8$ matrix for a given $\omega$. The left and right reservoirs are decoupled without tunneling, then $\mathcal{G}_0(\omega)$ is
\begin{equation}
\mathcal{G}_0(\omega)=\left(\begin{array}{cc}\mathcal{G}_{0L}(\omega) & 0 \\0 & \mathcal{G}_{0R}(\omega)\end{array}\right).
\end{equation}
Each $\mathcal G_{0j=\text{L},\text{R}}$ is a $4\times 4$ matrix of the form \cite{Cuevas, Bolech2004, Bolech2005}
\begin{equation}
\mathcal G_{0j}(\omega)=
\begin{bmatrix}
 \mathcal{G}^{R}_j(\omega)  & \mathcal{G}^{K}_j(\omega) \\
 0 & \mathcal{G}^{A}_j(\omega)
\end{bmatrix}, \label{eq:Greenfunction}
\end{equation}
where $\mathcal{G}^{R}_j(\omega)$, $\mathcal{G}^{A}_j(\omega)$ and $\mathcal{G}^{K}_j(\omega)$ are the retarded, advanced and Keldysh Green's functions, respectively.
Considering that each reservoir is in the thermal equilibrium, the Keldysh component of the Green's function can be obtained as $\mathcal{G}^{K}_j(\omega)=\tanh\left(\frac{\omega-\mu_j}{2T}\right)\left[\mathcal{G}_j^R(\omega)-\mathcal{G}_j^A(\omega)\right]$ at temperature $T$.
The mean-field Hamiltonian gives the retarded and the advanced Green's functions as \cite{Cuevas, Bolech2004, Bolech2005}
\begin{equation}\begin{aligned}
\mathcal{G}^{R(A)}(\omega)=&\frac{1}{\sqrt{\Delta_j^2-(\omega-\mu_j\pm i0^+)^2}}\times \\ &
\begin{bmatrix}
-(\omega-\mu_j\pm i0^+) & \Delta_j \\ \Delta_j & -(\omega-\mu_j\pm i0^+)
\end{bmatrix}.
\end{aligned}\end{equation}
It is straight-forward to calculate the inverse of the Green's function as
\begin{equation}
\mathcal G^{-1}_{0j}=
\begin{bmatrix}
 (\mathcal{G}^{R}_j )^{-1} & ({\mathcal{G}^{-1}_j})^{K} \\
 0 & (\mathcal{G}^{A}_j)^{-1}
\end{bmatrix}, \label{eq:Greenfunction}
\end{equation} where $({\mathcal{G}_j^{-1}})^{K}=-(\mathcal{G}^{R}_j )^{-1}\mathcal{G}_j^{K}(\mathcal{G}_j^{A})^{-1}$.

Thirdly, the tunneling between two reservoirs is modeled by local tunnelings as \cite{Cuevas, Bolech2004, Bolech2005}
\begin{equation}\begin{aligned}
&\mathcal{V}= \int_{-\infty}^{\infty}\frac{d\omega}{2\pi}  \sum_{\sigma,\alpha=1,2}  \\
&\left\{\mathcal{T}(\omega-V_g, V_D)\hat{\psi}^\dag_{L\alpha\sigma}(\omega-\mu_\text{L})\hat{\psi}_{R\alpha\sigma}(\omega-\mu_\text{R}) +\text{h.c.}\right\},
\end{aligned}\end{equation}
where the transmission amplitude $\mathcal{T}(\omega-V_g, V_D)$ is a function of $V_g$ and $V_D$ as discussed above. Thus, the full Green's function can be obtained by
\begin{equation}
\mathcal{G}=\left(\mathcal{G}^{-1}_0-\mathcal{V}\right)^{-1}.
\end{equation}
Although $\mathcal{G}^{-1}_0$ is diagnoal in the frequency bases, $\mathcal{V}$ introduces coupling between different frequencies. Hence, in practice, we discretize the frequency space and numerically calculate the inversion of the matrix.

Finally, it's convenient to calculate the current $I(t)=\frac{1}{2}\partial (N_R-N_L)/\partial t$ in the frequency space by introducing
\begin{equation}
I(t)=\int_{-\infty}^\infty\frac{d\Omega}{2\pi}I(\Omega)e^{-i\Omega t},
\end{equation}
and in the Keldysh formalism the current $I(\Omega)$ is written as
\begin{equation}\begin{aligned}
{I}(\Omega)=&\nonumber
-\frac{i}{2}\sum_{\sigma}\int_{-\infty}^\infty\frac{d\omega}{2\pi} \times\\
\Big\{\mathcal{T}&(\omega-V_g, V_D)
\langle\hat{\psi}_{R1\sigma}(\omega-\mu_\text{R}+\Omega)\hat{\psi}^\dag_{L2\sigma}(\omega-\mu_\text{L})\rangle\\
-\mathcal{T}&(\omega-V_g, V_D)\langle\hat{\psi}_{L1\sigma}(\omega-\mu_\text{L}+\Omega)\hat{\psi}^\dag_{R2\sigma}(\omega-\mu_\text{R})\rangle\Big\}.
\label{eq:current1}\end{aligned}\end{equation}
With the Green's function calculated above, it is straightforward to obtain $I(\Omega)$. Here the frequency $\Omega$ can only take a series of discrete values as $\Omega_m=2m\delta\mu$ with $m=0,\pm1,\pm2,...$ \cite{Cuevas}, where $\delta\mu=\mu_\text{L}-\mu_\text{R}$ is the bias voltage. Denoting $I(\Omega_m)\equiv I_m$, the total current for a fixed bias voltage can be written as
\begin{equation}\begin{aligned}
I(t)=I_0+2\sum_{m=1}^\infty \big[{\rm Re}(I_m)\cos(\Omega_mt)+{\rm Im}(I_m)\sin(\Omega_mt)\big].\label{eq:current2}
\end{aligned}
\end{equation}

\begin{figure}[t]
\includegraphics[width=0.48\textwidth]{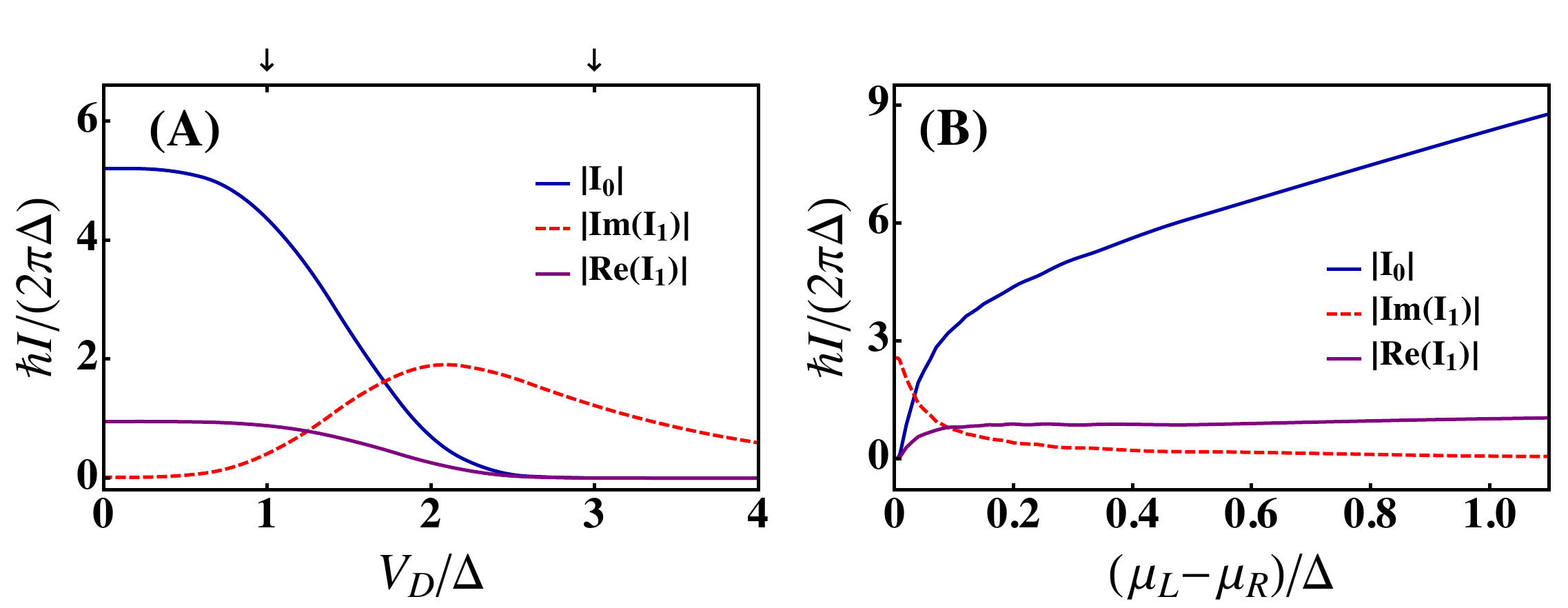}
\caption{(Color online) $|I_0|$, and the real and imaginary part of $I_1$ as a function of $V_\text{D}/\Delta$ for fixed $\delta\mu=0.2\Delta$ (A), and as a function of $\delta\mu/\Delta$ for fixed $V_\text{D}=\Delta$ (B). Here we have fixed $V_\text{g}=-14\Delta$. }
\label{fig:current}
\end{figure}

\textit{Controlled Transport.} In Fig. \ref{fig:current}, we first show the current $I_0$ and $I_1$ as functions of $V_\text{D}/\Delta$ and $\delta\mu/\Delta$. We have verified that the higher component currents are much smaller than these two and can be safely ignored. The $I_0$ part is the dc component, which is resulted from the quasi-particle transport by the multiple Andreev reflections. The components of ${\rm Re}(I_1)\cos(\Omega_1t)$ and ${\rm Im}(I_1)\sin(\Omega_1t)$ are the ac parts and usually referred as the ``cosine" term and ``sine" terms. They describe the phase coherent transport of the Cooper pairs \cite{Barone}. The sine term ${\rm Im}(I_1)\sin(\Omega_1t)$ is related to the usual Josephson current. The cosine term was also predicted by Josephson \cite{Josephson} originally. It was first observed in the experiment by Pederson, Finnegan, and Langenberg \cite{Pederson}. In the usual weak link discussion, the cosine term vanishes in the first order of perturbation theory, and presents at the second order calculation.

Because the dc current is generated by the multiple Andreev reflection, if $n$ is the smallest integer such that $n\delta \mu>2\Delta$, it takes at least $(2n-1)$-step of tunnelings in order to generate the dc current, and therefore the current is proportional to $\mathcal{T}^{2n}$ \cite{Cuevas,Bolech2005}. Thus, the dc current is dominative either when $V_\text{D}$ is small and $\mathcal{T}$ is close to unity, as shown in Fig. \ref{fig:current}(A), or when $\delta\mu$ is larger comparing to $\Delta$ and $n$ is small, as shown in Fig. \ref{fig:current}(B). In another word, either when $V_\text{D}$ increases and $\mathcal{T}$ decreases, or when $\delta\mu$ becomes small, the dc current gradually decreases until vanishing. In these two regimes, the Josephson current will become dominative, as one can also see from Fig. \ref{fig:current}.

\begin{figure}[t]
\includegraphics[width=0.48\textwidth]{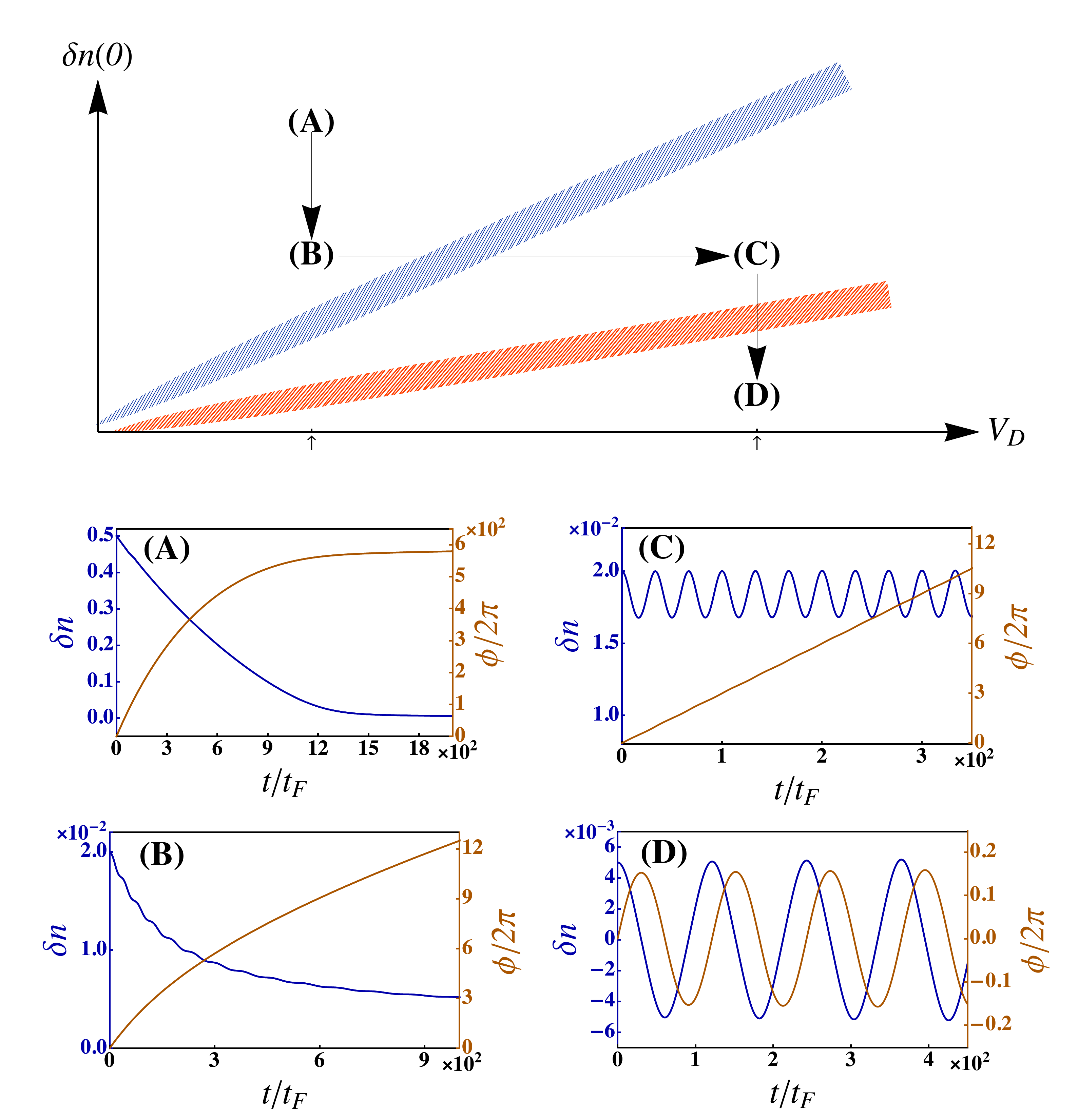}
\caption{(Color online) Schematic of different transport regimes with different initial atom number difference $\delta n$ and different $V_\text{D}$. The two arrows on the axes of $V_D$ indicate the values we take for graph A-D, which correspond to the values labeled by arrows in Fig. \ref{fig:current} (A). (A-D) represent four typical dynamical behaviors for $\delta n$ and $\phi$. For (A), $\delta n(0)=0.5$, $V_\text{D}=\Delta$ and corresponding $\mathcal{T}\sim 0.9$. (B) has the same $V_\text{D}$ and $\mathcal{T}$ as (A), but smaller initial $\delta n(0)=0.02$. For (C), $\delta n(0)=0.02$, $V_\text{D}=3\Delta$ and corresponding $\mathcal{T}\sim 0.5$. (D) has the same $V_\text{D}$ and $\mathcal{T}$ as (C), but different initial $\delta n(0)=0.005$.    }
\label{fig:flow}
\end{figure}

In realistic cold atom experiments, instead of studying transport with a fixed bias voltage as in condensed matter system, one starts with an initial atom number imbalance and monitors how this imbalance evolves as a function of time. To investigate this dynamics, we employ a coupled dynamical equations as follows
\begin{equation}\begin{aligned}
\frac{d\delta n}{dt}&=-I_0-2{\rm Re}(I_1)\cos\phi-2{\rm Im}(I_1)\sin\phi\\
\frac{d\phi}{dt}&=2\delta\mu(t),
\label{EqNt}
\end{aligned}\end{equation}
where $\phi$ is the phase difference between the two reservoirs, and $\delta\mu$ is related to the atom imbalance by
\begin{equation}
\frac{\mu_L}{\mu_R}=\left(\frac{1+\delta n}{1-\delta n}\right)^{2/3}
\end{equation}
and $\delta n=(N_L-N_R)/(N_L+N_R)$. Here the major assumption is that the tunneling time is much longer than the local equilibrium time of the reservoir (characterized by $t_\text{F}=2\pi\hbar/E_\text{F}$), such that we can apply above results with fixed $\delta\mu$ to any instantaneous time. From the results we obtained (as shown in Fig. \ref{fig:flow}), this assumption is indeed well obeyed.

The different regimes of tunneling dynamics is summarized in Fig. \ref{fig:flow} with two tunable parameters $\delta n$ and $V_\text{D}$. Let us analyze the evolution between different regimes as follows:

\textbf{A $\rightarrow$ B}: In case (A), $I_0$ is the most dominative component and $\delta n$ quickly drops to zero. From case (A) to case (B), the initial $\delta n$ decreases. Because $I_0$ is less dominative in the small $\delta \mu$ regime, as shown in Fig. \ref{fig:flow}(B), one can see that a small oscillation of $\delta n$ becomes visible, and at mean time the long time saturation value of $\delta n$ is a finite value instead of dropping to zero.

\textbf{B $\rightarrow$ C}: From case (B) to case (C), the initial $\delta n$ is about the same, but $V_\text{D}$ increases and $\mathcal{T}$ decreases. The Josephson effect gradually becomes dominant over the multiple Andreev reflection. This crossover happens in the way that on one hand, the saturation value of $\delta n$ increases until eventually the drop of $\delta n$ becomes insignificant, and on the other hand, the oscillation becomes more profound. As a result, $\delta n$ oscillates around a finite value. This is known as the ``self-trapping" regime in the previous study of the Josephson oscillations in a Bose-Einstein condensate of bosons \cite{Smerzi}.

\textbf{C $\rightarrow$ D}: In both cases of (C) and (D), $V_\text{D}$ and $\mathcal{T}$ are fixed in the regime where $\text{Im}(I_1)$ is the most dominative component. The only difference is that the initial $\delta n$ in case (D) is much smaller than that of case (C). Only keeping the $\sin\phi$ term in the first equation of Eq. \ref{EqNt}, it is known that Eq. \ref{EqNt} can be mapped to the dynamical equation of a classical pendulum, where two different solutions can be found depending on the initial conditions. The first is an oscillation around the global minimum when $\delta n$ is small, corresponding to the conventional Josephson effect, and the other is a continuous clockwise (or anti-clockwise) rotation when $\delta n$ is large, corresponding to the self-trapping \cite{Smerzi}. Similar crossover from Josephson to self-trapping regime has also been observed previously in Bose condensate of bosons \cite{Albiez}.

%-----------------------figure--------------------------------------------------
\begin{figure}[t]
\includegraphics[width=0.48\textwidth]{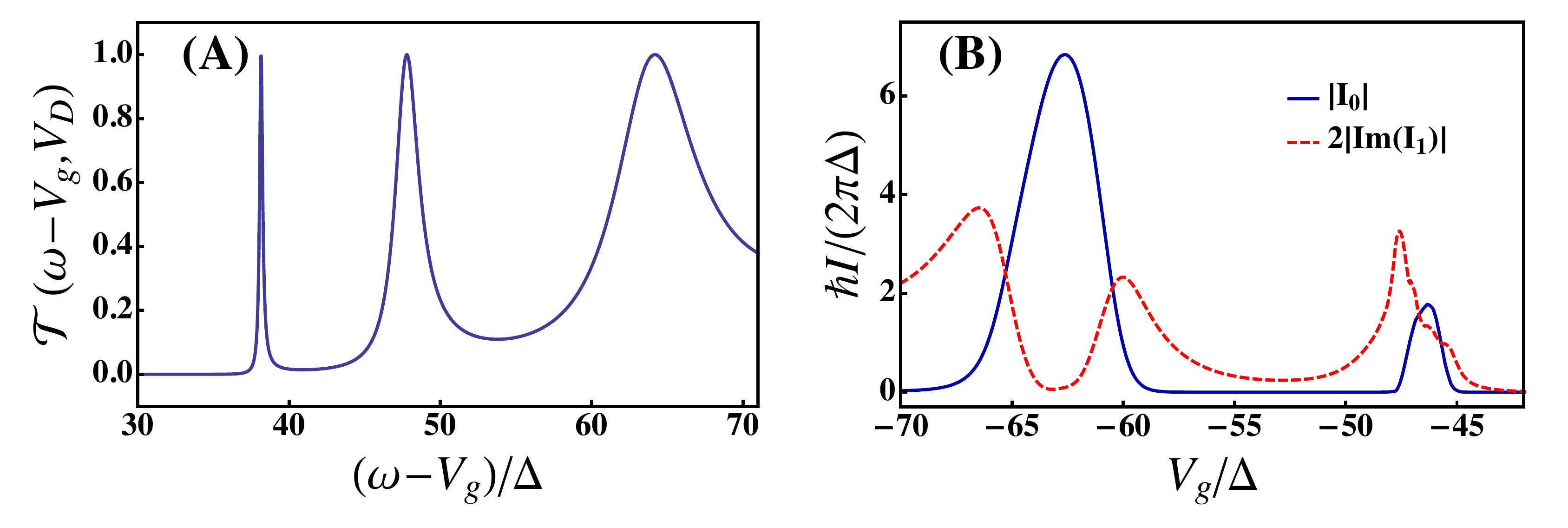}
\caption{(Color online)(A) The transmission amplitude $\mathcal T(\omega-V_g, V_D)$ for the case of two delta-function potentials with $V_D=8\Delta$. (B) The currents $I_0$ and ${\rm Im}(I_1)$ as functions of the gate potential for a fixed chemical potential bias $\delta\mu=0.6\Delta$.}
\label{fig:twoDMD}
\end{figure}
%-----------------------figure--------------------------------------------------

\textit{Tuning Transport with Gate Potential.} For a single delta-function potential barrier, $\mathcal{T}$ is a monotonic function of $\omega-V_\text{g}$. And for multiple delta-function potentials, due to the interference effect, $\mathcal{T}$ as a function of $\omega-V_\text{g}$ exhibits much richer structure, as shown in Fig. \ref{fig:twoDMD}(A). Therefore, by varying $V_\text{g}$, $\mathcal{T}$ at the fixed Fermi surface can vary significantly, and consequently, it leads to a large variation of the relative strengthes between $I_0$ and ${\rm Im}(I_1)$, as shown in Fig. \ref{fig:twoDMD}(B). Thus, in this case, the different regimes of transport behaviors discussed above can also be controlled by $V_\text{g}$ instead of $V_\text{D}$.

\textit{Outlook.} In summary, we have presented a system that three distinct behaviors of transport between two Fermi superfluids can all be realized, and the crossover between them can be tuned by both the initial atom number imbalance and the transmission coefficient. Given that the experimental technique required has been reported, our prediction can now be directly applied to current experimental setup. Further generalization of this study can also study spin and heat transport with tunable transmission coefficient through a quantum point contact.

\textit{Acknowledgements.}
We would like to thank Shizhong Zhang and Shun Uchino for discussions. This work is supported by MOST under Grant No. 2016YFA0301600 and NSFC Grant No. 11734010.

\end{document}